\documentclass[a4paper,11pt]{article}
\usepackage{pos}
\usepackage{multirow,multicol}
\usepackage{hyperref}
\usepackage{blindtext}
\usepackage{amsmath}
\usepackage{amsfonts}
\usepackage{graphicx}
\usepackage{caption,xcolor}
\usepackage{subfigure}
\usepackage{mathtools}

\newcommand{\nn}{\nonumber}
\newcommand{\lsim}{\mathrel{\mathop{\kern 0pt \rlap
			{\raise.2ex\hbox{$<$}}}
		\lower.9ex\hbox{\kern-.190em $\sim$}}}
\newcommand{\gsim}{\mathrel{\mathop{\kern 0pt \rlap
			{\raise.2ex\hbox{$>$}}}
		\lower.9ex\hbox{\kern-.190em $\sim$}}}

\newcommand{\mpm}{\mu^\pm}
\newcommand{\mpmm}{\mu^+\mu^-}

\title{(New) Physics at a multi-TeV $\mu$ Collider}

\author*[a]{Antonio Costantini}

\affiliation[a]{Centre for Cosmology, Particle Physics and Phenomenology {\rm (CP3)},\\
Universit\'e Catholique de Louvain, Chemin du Cyclotron, B-1348 Louvain la Neuve, Belgium}

\emailAdd{antonio.costantini@uclouvain.be}

\abstract{We present the results of the physics reach of a multi-TeV muon collider for popular models of physics beyond the Standard Model. We include also details about a model predicting the scalar dark matter in the spectrum. Finally we present some preliminary results about the Effective Vector Approximation and its implementation in the Monte Carlo generator {\sc MadGraph5\_aMC@NLO}.}

\FullConference{%
  *** The European Physical Society Conference on High Energy Physics (EPS-HEP2021), ***\\
  *** 26-30 July 2021 ***\\
  *** Online conference, jointly organized by Universität Hamburg and the research center DESY ***
}

\begin{document}
\maketitle

\section{Introduction}

Standing out among the important results that the Large Hadron Collider (LHC) has thus far delivered  are the discovery of the Higgs boson $(H)$ and the measurements of its properties. 
On the other hand, long-awaited evidence of new physics based on theoretical arguments, such as the stabilization of the electroweak (EW) scale, 
or on experimental grounds, such as dark matter and neutrino masses, have evaded our scrutiny.  Depending on the properties of the new phenomena,  either ``low-energy'' precision measurements or searching for new states in ``high-energy'' direct production may be the most sensitive and informative strategy to follow.
 Accelerating muons would allow one to merge the best of both hadron and $e^+e^-$ colliders i.e., a high energy reach on one side and a ``clean'' environment on the other.

Our starting point is the observation that at sufficiently high energies  the EW vector boson fusion (EW VBF) become the dominant production mode at a multi-TeV lepton collider.
We anticipate this holding for all Standard Model (SM) final states relevant to studying  the EW sector and/or the direct search of (not too heavy) new physics. 

Our main interest will be discussing the Beyond-the-Standard-Model (BSM) capabilities of a multi-TeV muon collider and present some preliminary results concerning the  Effective Vector Boson Approximation (EVA) and its implementation in Monte Carlo generators. Nonetheless we present as a first results some of the relevant production processes via VBF, namely
\begin{equation}
\mu^+\mu^-\to X\, {\nu}_{\mu}\overline{\nu}_{\mu}
\end{equation}
 and their counterpart from muon annihilation (through $s$-channel) in the SM in Table~\ref{tab:neutralSM}. The VBF processes shown here and also their partner with muons in the final state ($\mu^+\mu^-\to X\, \mu^+\mu^-$) exhibit a growth in the cross-section with energy, and eventually overcome the $s$-channel production processes \cite{Costantini:2020stv}.

\begin{table}[!t]
\begin{center}
\resizebox{\textwidth}{!}{
\begin{tabular}{l||cc||cc||cc||cc}
\multirow{2}{*}{$\sigma$ [fb]} &        \multicolumn{2}{c}{$\sqrt s=$ 1 TeV} &    \multicolumn{2}{c}{$\sqrt s=$ 3 TeV} &    \multicolumn{2}{c}{$\sqrt s=$ 14 TeV} &      \multicolumn{2}{c}{$\sqrt s=$ 30 TeV}\\
		&VBF&s-ch.&VBF&s-ch.&VBF&s-ch.&VBF&s-ch\\

$t \bar{t}$           	&  4.3$\cdot 10^{-1}$ &1.7$\cdot 10^{2}$ &  5.1$\cdot 10^{0}$ &1.9$\cdot 10^{1}$ &  2.1$\cdot 10^{1}$ &8.8$\cdot 10^{-1}$ &  3.1$\cdot 10^{1}$ &1.9$\cdot 10^{-1}$ \\
\hline
$H$                   	&        2.1$\cdot 10^{2}$ &- &        5.0$\cdot 10^{2}$ &- &        9.4$\cdot 10^{2}$ &- &        1.2$\cdot 10^{3}$ &- \\
$H H$                 	&        7.4$\cdot 10^{-2}$ &- &        8.2$\cdot 10^{-1}$ &- &        4.4$\cdot 10^{0}$ &- &        7.4$\cdot 10^{0}$ &- \\
\hline
$W W$		&1.6$\cdot 10^{1}$&2.7$\cdot 10^{3}$&1.2$\cdot 10^{2}$&4.7$\cdot 10^{2}$&5.3$\cdot 10^{2}$&3.2$\cdot 10^{1}$&8.5$\cdot 10^{2}$&8.3$\cdot 10^{0}$\\
$Z Z$		&6.4$\cdot 10^{0}$&1.5$\cdot 10^{2}$&5.6$\cdot 10^{1}$&2.6$\cdot 10^{1}$&2.6$\cdot 10^{2}$&1.8$\cdot 10^{0}$&4.2$\cdot 10^{2}$&4.6$\cdot 10^{-1}$\\

\end{tabular}
}
\caption{$W^+W^-$ fusion and analogous $s$-channel annihilation cross sections $\sigma$ [fb] for various processes in the SM as a function of collider energy $\sqrt{s}$ [TeV].}
\label{tab:neutralSM}
\end{center}
\end{table}

\section{New Physics Potential at a multi-TeV $\mu$ Collider}\label{sec:bsmMu}

In this section, we present a survey of  BSM models and the potential sensitivity of a $\mpmm$ collider. 
Explicitly, we consider the $s$-channel annihilation and VBF processes
\begin{equation}
\mpmm ~\to~ X \quad\text{and}\quad
\mpmm ~\to~ X \ell \ell'.
\end{equation}
Here, $\ell\in\{\mpm,\overset{(-)}{\nu_\mu}\}$ and $X$ is some BSM final state, which may also include SM particles.
While via $s$-channel annihilation one can accesses to the highest available c.m.~energies, 
there's a suppression in the cross-section that scales as $\sigma\sim1/s$ when far above production threshold.
On the other hand, in VBF,  the emission of transversely polarized, $t$-channel bosons gives rise to logarithmic factors that grow with the available collider energy.

We can roughly estimate the collider energy $\sqrt{s}$ at which $\sigma^{\rm VBF}$ surpasses $\sigma^{s-ch.}$ for a given final-state mass $M_X$.
Essentially, one must solve for when
\begin{equation}
\frac{\sigma^{\rm VBF} }{\sigma^{s-ch.}} \sim 
\mathcal{S} \left(\frac{g_W^2}{4\pi}\right)^2   \left(\frac{s}{M_X^2}\right) \log^2\frac{s}{M_V^2}\log\frac{s}{M_X^2} > 1.
\label{eq:bsm_vbf_scaling}
\end{equation}

\begin{figure}[t!]
\centering\subfigure[]{\includegraphics[width=.58\textwidth]{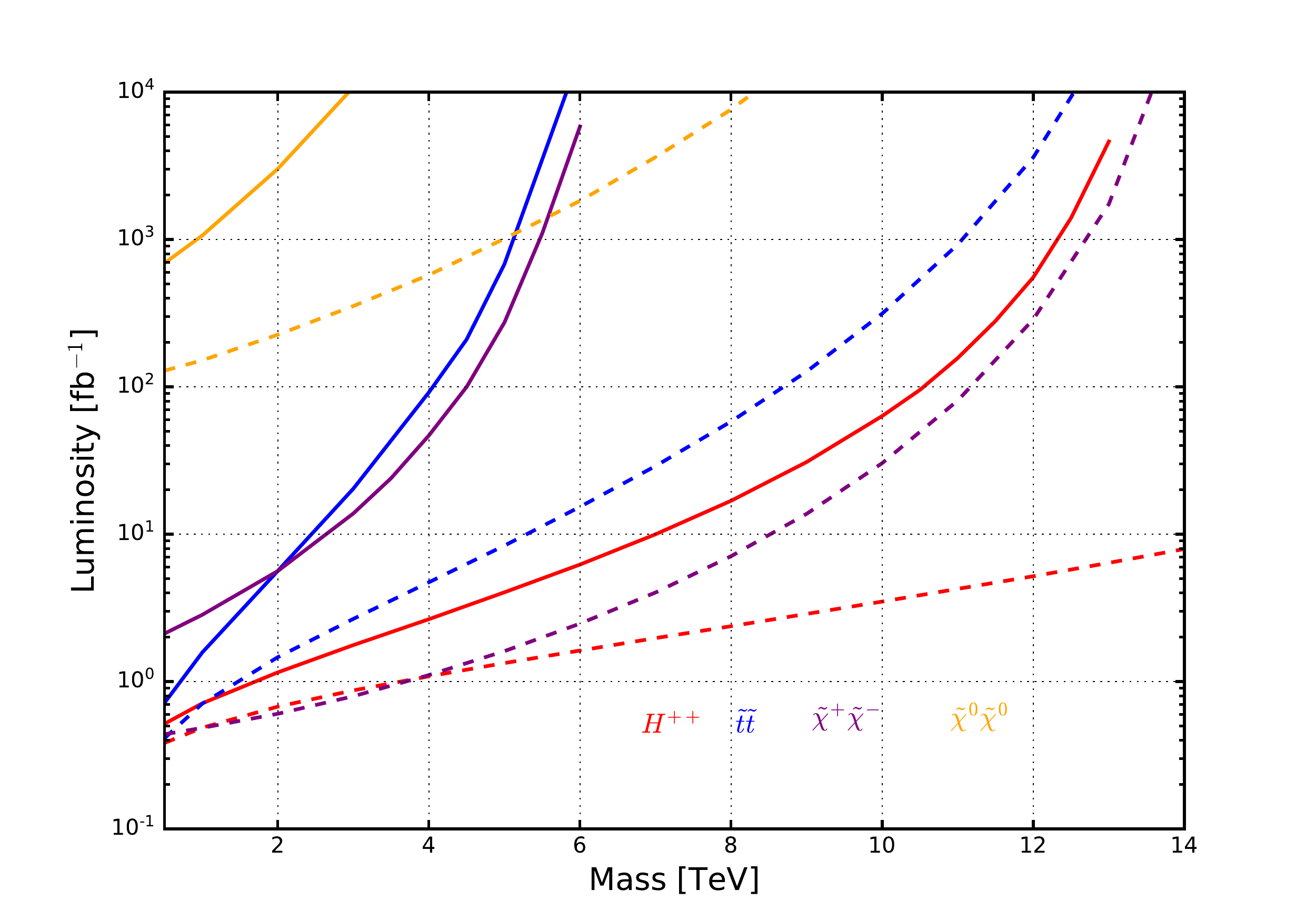}}
	\subfigure[]{\includegraphics[scale=.3]{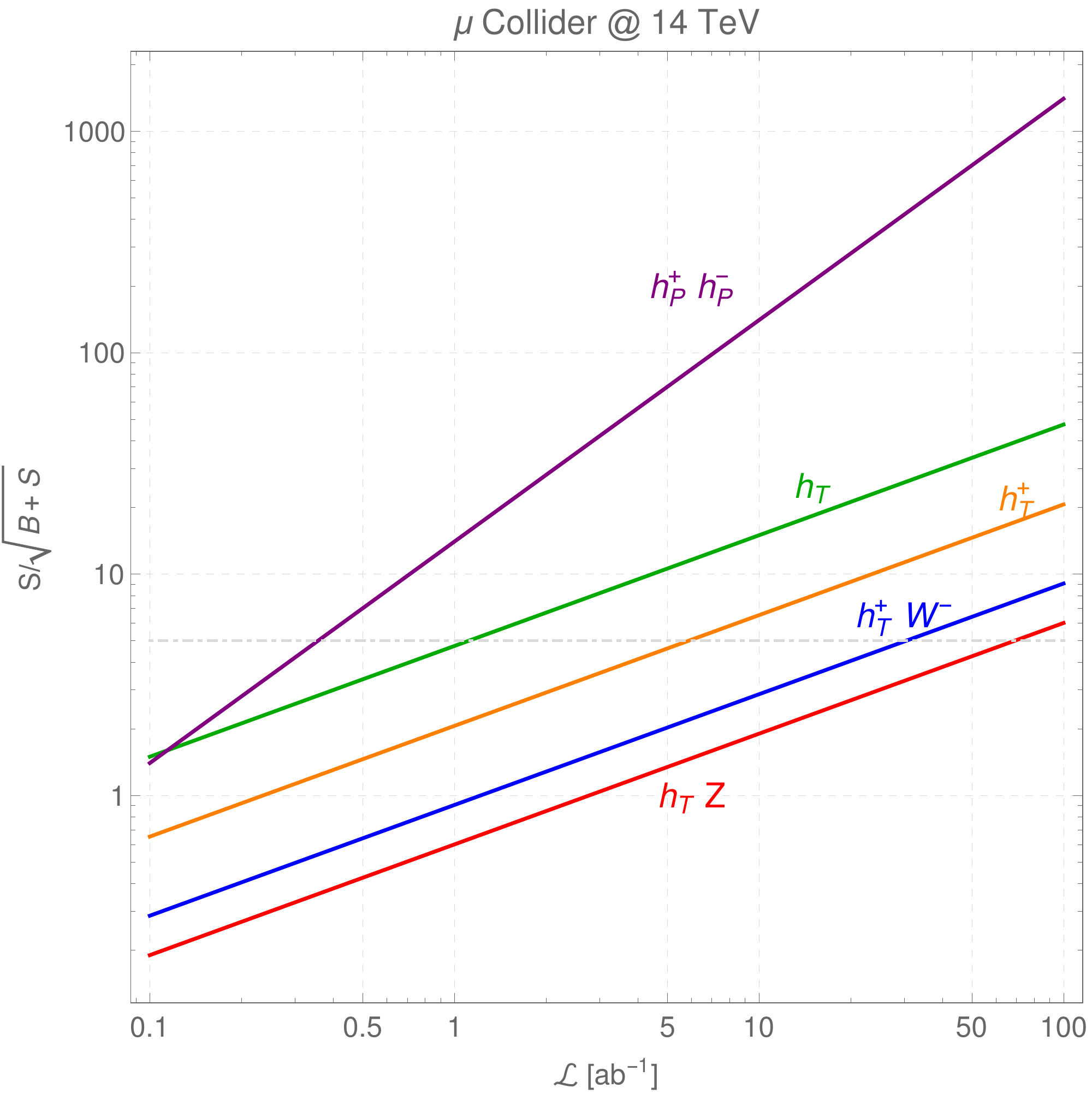}}
\caption{
(a) Required luminosity [fb] for a  $5\,\sigma$ discovery of 
$H^{++}$ (red) in the GM model;
$\tilde{t}\overline{\tilde{t}}$ (blue), 
$\tilde{\chi}^+\tilde{\chi}^-$ (purple), and
$\tilde{\chi}^0\tilde{\chi}^0$ (yellow)  from in the MSSM,
using VBF in $\sqrt{s}=14$ TeV (solid) and $30$ TeV (dashed)  muon collisions. (b) Significance vs luminosity for the $h_T$, $h_T^\pm$, $h_T^+\, W^-$, $h_T\, Z$ and $h_P^+ h_P^-$ production processes through VBF at a 14 TeV muon collider.
}
\label{fig:sensi}
\end{figure}

Finally we investigated the sensitivity of EW VBS to a variety of BSM scenarios at multi-TeV muon colliders.
In order to give an overview picture of this reach, we present in figure~\ref{fig:sensi} the
requisite integrated luminosity $\mathcal{L}$ [fb$^{-1}$] for a $5\sigma$ discovery as a function of new particle mass in $\sqrt{s}=14$ TeV (solid) and $30$ TeV (dashed) muon collisions.
We consider specifically 
the doubly charged Higgs $H^{++}$ (red) from the GM model, $\tilde{t}\overline{\tilde{t}}$ (blue), $\tilde{\chi}^+\tilde{\chi}^-$ (purple), and $\tilde{\chi}^0\tilde{\chi}^0$ (yellow) pairs from the MSSM.
As dedicated signal and background analyses are beyond the scope of this document, we crudely assume a zero background hypothesis and full signal acceptance.
We therefore also use
as a simple measure of statistical significance $(\mathcal{S})$ the formula,  $\mathcal{S}=\sqrt{\mathcal{L}\times\sigma}$.

Let us now consider an extension of the Standard Model with a complex triplet with $Y=0$, which we name complex Triplet extension of the Standard Model (cTSM). The gauge and fermion sectors are identical to the Standard Model ones, whereas the scalars of the model are
\begin{eqnarray}
\Phi=\left(
\begin{array}{c}
\phi^+\\
\Phi_0
\end{array}
\right),
\qquad
T=\frac{1}{\sqrt2}\left(
\begin{array}{cc}
t_0&\sqrt2\, t_1^+\\
\sqrt2\, t_2^-&-t_0
\end{array}
\right).
\end{eqnarray}
Here we will describe the main features of the model and refer to \cite{Bandyopadhyay:2020otm} for the details of the analysis. As already stated, the only difference with the SM relies in the scalar sector. Apart from the SM(-like) Higgs boson we have a massive pseudoscalar and two massive charged scalars, namely $h_D, h_T, a_P, h^\pm_{T,P}$. Due to symmetry reasons, the pseduscalar and one of the charged Higgs bosons are pure even after electroweak symmetry breaking (EWSB). The most important consequence is that the massive pseudoscalar is a natural dark matter candidate \cite{Bandyopadhyay:2020otm}.

We have considered the physics reach for some processes at a muon collider \cite{Bandyopadhyay:2020otm}. These results are presented in figure~\ref{fig:sensi} (b). We plot the significance as function of the luminosity for the $h_T$, $h_T^\pm$, $h_T^+\, W^-$, $h_T\, Z$ and $h_P^+h_P^-$ production processes through VBF at a 14 TeV muon collider. In the definition of the significance, $\sigma=S/\sqrt{S+B}$, $S$ and $B$ stand for the number of events for the signal and the background respectively, 
\begin{align}
S:& \;\sigma(\mu^+\mu^-\,\to h_T \;\nu_\mu\bar\nu_\mu) \times Br(h_T\to W^+W^-) \cdot \mathcal{L},\\
B:& \;\sigma(\mu^+\mu^-\,\to W^+W^- \;\nu_\mu\bar\nu_\mu) \cdot \mathcal{L},
\end{align}
with $M(W^+ W^-)=m_{h_T}\pm5$ GeV. A similar strategy is applied to the single production of $h_T^\pm$ and the pair-production $h_T Z$ and $h_T^+W^-$
. 
whereas the pair production $h_P^+h_P^-$ has been considered background-free. The pure charged triplet $h_P^\pm$ has a single decay channel, namely $h_P^+\to a_P (W^+)^*$. Whereas the pseudoscalar is undetectable, the process will give rise to displaced off-shell $W$ boons: there is no SM process that have this particular final state.  This also gives rise to displaced leptons/jets plus missing energy in the final-state.

\section{Effective Vector Approximation}
In this section, we summarize the Effective Vector Boson Approximation (EVA) and its use in evaluating scattering cross sections in many-TeV $\mu^+\mu^-$ collisions, in the framework of the Monte Carlo generator {\sc MadGraph5\_aMC@NLO} \cite{Alwall:2014hca,BuarqueFranzosi:2019boy,Ruiz:2021tdt}. We focus on muons but the EVA is, in principal, applicable to other colliders. However, this may require convolutions with additional PDFs~\cite{Dawson:1984gx}.

\begin{figure}[t!]
\centering
\subfigure[]{\includegraphics[width=.4\textwidth]{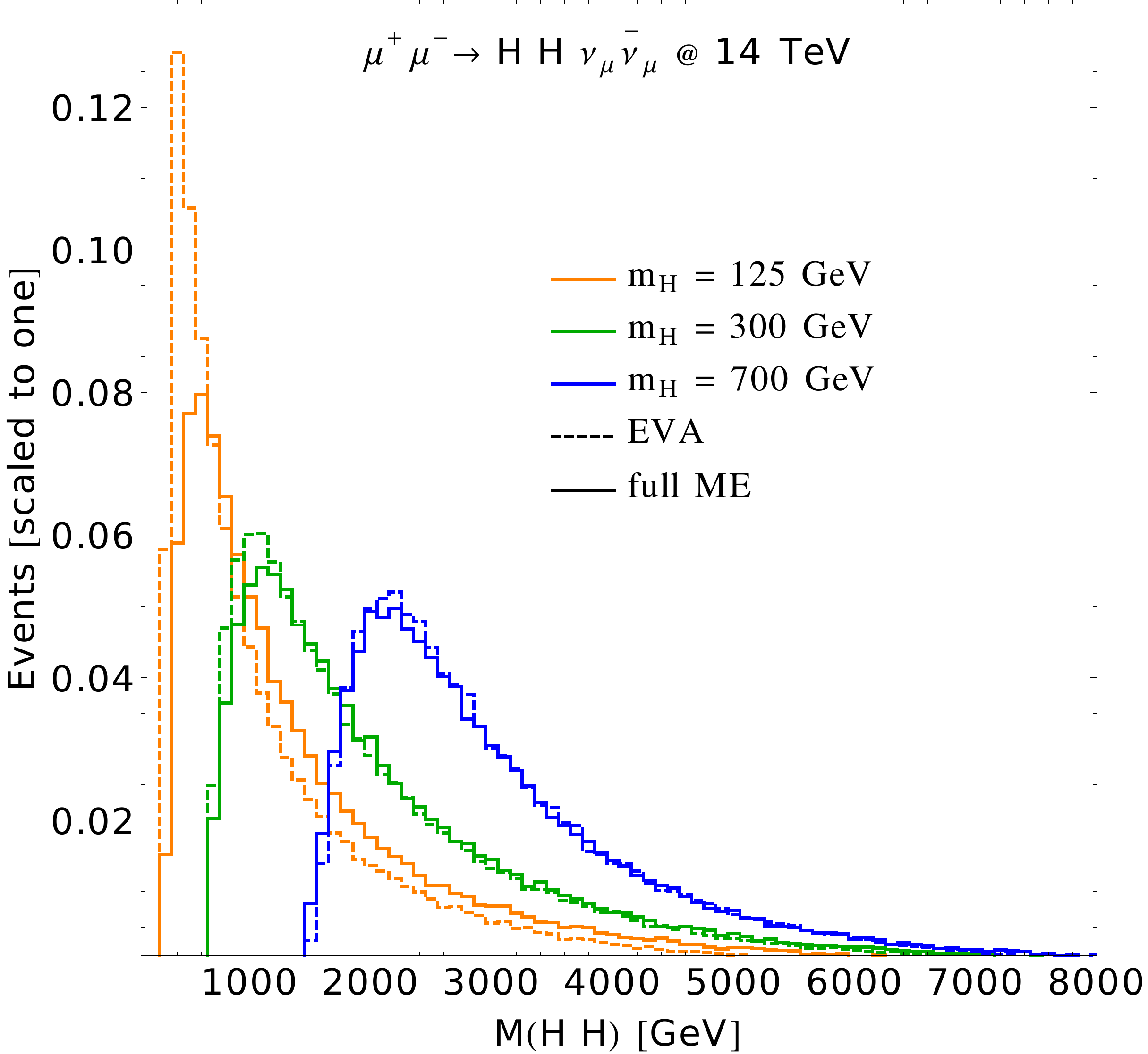}}\hspace{.2cm}
\subfigure[]{\includegraphics[width=.38\textwidth]{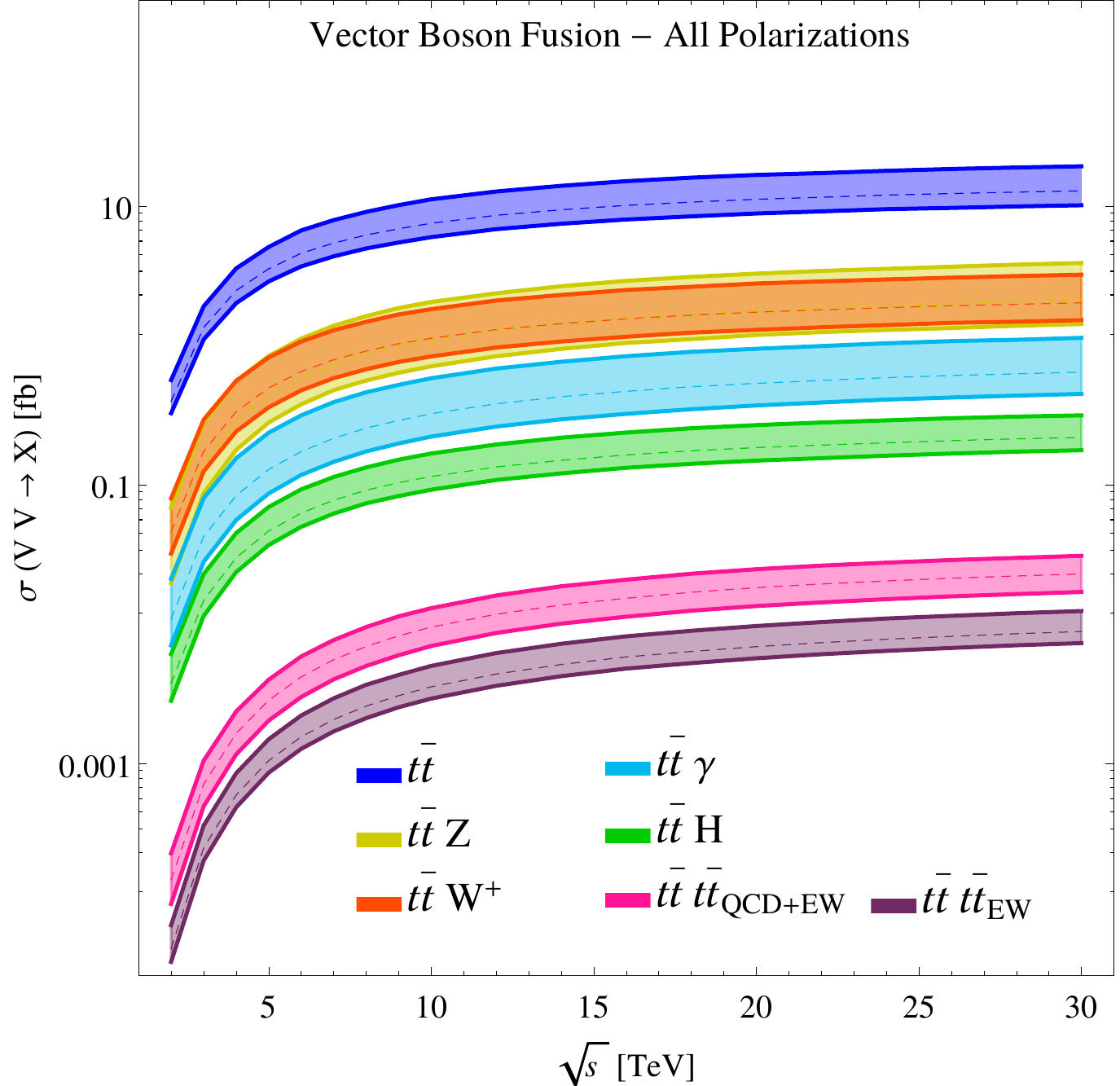}}
\caption{
(a) Invariant mass distribution for the Higgs pair production at a 14 TeV $\mu$ Collider, in the EVA approximation (dashed lines) and in the full ME computation (solid lines) for different values of $m_H$.  (b) Fiducial cross-section under the EVA for $t\bar t + X$ associated production in $\mu^+\mu^-$ collisions coming from all initial-state EW bosons polarizations.
}
\label{fig:ttX}
\end{figure}

We described the production of an $n$-body, final state $\mathcal{F}$ in $\mu^+\mu^-$ collisions at  an energy of $\sqrt{s}$ via high-energy VBF process $ V_{\lambda_A}V'_{\lambda_B} \to \mathcal{F}$ with EVA. In practice, this means working from a factorization theorem given by
\small
\begin{align}
\sigma(\mu^+ \mu^- &\to \mathcal{F} + X)=\nn\\
&\frac{1}{1+\delta_{VV'}}
\sum_{V_{\lambda_A},V'_{\lambda_B}}
\int dPS_{n} \times
\left[
f_{V_{\lambda_A}/\mu^+}
f_{V'_{\lambda_B/\mu^-}}
+
f_{V'_{\lambda_B}/\mu^+}
f_{V_{\lambda_A/\mu^-}}
\right]\times \frac{d\hat{\sigma}(V_{\lambda_A}V'_{\lambda_B} \to \mathcal{F})}{dPS_{n}}.
\label{eq:factTheorem}
\end{align}
\normalsize
Here, $\sigma$ is the (beam-level) inclusive cross section for the production of $\mathcal{F}$ in association with an arbitrary state $X$. Explicitly, $X$ consists of at least two leptons $l$, where $l=\mu^\pm, \nu_\mu,$ or $\overline{\nu_\mu}$.
The summation runs over all polarized EW boson $V_{\lambda} \in \{W^\pm_\lambda, Z_\lambda, \gamma_\lambda\}$,  with $\lambda\in\{0,\pm1\}$.

The polarized weak boson PDFs to $\mathcal{O}(g_W^2)$ for LH $(f_L)$ and RH $(f_R)$ fermions are \cite{Ruiz:2021tdt,Borel:2012by,Dawson:1984gx}
\begin{subequations}
\label{eq:formalism_pdfs_q2}
\begin{align}
f_{V_+/f_L}(\xi,\mu_f^2) 	&= \frac{g_V^2}{4\pi^2} \frac{g_L^2(1-\xi)^2}{2\xi} \log \left[\frac{\mu_f^2}{M_V^2}\right], \quad f_{V_+/f_R}=\left(\frac{g_R}{g_L}\right)^2 \times f_{V_-/f_L}, \\
f_{V_-/f_L}(\xi,\mu_f^2) 	&= \frac{g_V^2}{4\pi^2} \frac{g_L^2}{2\xi} \log \left[\frac{\mu_f^2}{M_V^2}\right], \quad f_{V_-/f_R}= \left(\frac{g_R}{g_L}\right)^2 \times f_{V_+/f_L},\\
f_{V_0 / f_L}(\xi,\mu_f^2) 	&= \frac{g_V^2}{4\pi^2} \frac{g_L^2(1-\xi)}{\xi},\quad f_{V_0/f_R}= \left(\frac{g_R}{g_L}\right)^2 \times f_{V_0/f_L}.  
\end{align}
\end{subequations}

We summarize in figure \ref{fig:ttX} some preliminary results obtained for the study of high energetic muon collision. The plot in figure \ref{fig:ttX} (a) shows the invariant mass distribution (scaled to one) of a Higgs pair, produced via VBF at a $\mu$ Collider running at $\sqrt s=14$ TeV. The computation with the full Matrix Element (ME) process is shown in solid lines ($\mu^+ \mu^-\to HH\, \nu_\mu \bar\nu_\mu$) whereas dashed lines depict the EVA counterpart of the process ($W^+W^- \to HH$). Orange, green and blue line correspond to $m_H= 125, 300, 700$ GeV respectively. The accordance between the full ME computation and the EVA approximation is clearly better for $m_H\gsim300$ GeV \cite{Ruiz:2021tdt}. Figure \ref{fig:ttX} (b) shows the cross-section as function of $\sqrt s$ for the various processes $\sum_{V_{\lambda_A},V'_{\lambda_B}}V_{\lambda_A}V'_{\lambda_B}\to t\bar t + X$, $X$ particle(s) being a SM boson or another top pair. As for the cuts applied to such processes, we have $\textrm M(t\bar t X) > 1$ TeV. One of the main feature of this plot is the clear hierarchy of the cross sections and the fact that $\sigma_{t\bar t t \bar t_{\mathrm{ QCD+EW}}}\gsim\sigma_{t\bar t t \bar t_{\mathrm{EW}}}$, i.e. the cross-section of four tops involving only colorless intermediate states ($\sigma_{t\bar t t \bar t_{\mathrm{EW}}}$) is slightly lower than the corresponding process where also gluons are involved ($\sigma_{t\bar t t \bar t_{\mathrm{QCD+EW}}}$) \cite{Ruiz:2021tdt}.

\section{Conclusions}\label{sec:conc}
We explored a variety of simplified extensions of the SM and shown  how large VBF luminosities can maximize the direct search for new physics. This feature is similar in all the BSM models considered and constitutes a solid motivation for a multi-TeV $\mu$ Collider, together of course with the achievement for SM and SMEFT that a $\mu$ Collider can give.
We also consider in some detail a specific model with a scalar dark matter candidate showing the interplay between cosmological constraint and collider searches at a multi-TeV $\mu$ Collider. Finally, we've presented the EVA approximation and its implementation in the framework of the Monte Carlo generator {\sc MadGraph5\_aMC@NLO}, which will be very soon available.

\end{document}